\documentclass[twocolumn,prl,showpacs,superscriptaddress,showpacs]{revtex4-1}
\usepackage{amssymb,amsmath}
\usepackage{graphicx}
\usepackage{bm}
\usepackage{dcolumn}
\usepackage{float}
\usepackage{url}
\usepackage{color}
\usepackage{subfigure}
\usepackage{siunitx}
\usepackage{txfontsb}
\usepackage[OT1]{fontenc} 
\usepackage[driverfallback=dvipdfm]{hyperref}
\hypersetup{pdfpagemode=FullScreen,colorlinks=true,breaklinks,urlcolor=blue,linkcolor=blue,citecolor=blue}

\begin{document}

\title{Observation of the Hopf Links and Hopf Fibration in a 2D Topological Raman Lattice}

\author{Chang-Rui Yi}
\affiliation{Hefei National Laboratory for Physical Sciences at Microscale
and Department of Modern Physics, University of Science and Technology of China, Hefei, Anhui 230026, China}
\affiliation{Shanghai Branch, CAS Center for Excellence and Synergetic Innovation Center in Quantum Information and Quantum Physics, University of Science and Technology of China, Shanghai 201315, China}

\author{Jin-Long Yu}
\affiliation{Institute for Advanced Study, Tsinghua University, Beijing 100084, China}
\affiliation{Institute for Quantum Optics and Quantum Information of the Austrian Academy of Sciences, Innsbruck A-6020, Austria}

\author{Wei Sun}
\author{Xiao-Tian Xu}
\author{Shuai Chen}
\email{shuai@ustc.edu.cn}
\author{Jian-Wei Pan}
\email{pan@ustc.edu.cn}
\affiliation{Hefei National Laboratory for Physical Sciences at Microscale
and Department of Modern Physics, University of Science and Technology of China, Hefei, Anhui 230026, China}
\affiliation{Shanghai Branch, CAS Center for Excellence and Synergetic Innovation Center in Quantum Information and Quantum Physics, University of Science and Technology of China, Shanghai 201315, China}

\date{\today}

\begin{abstract}
A dynamical Hopf insulator is experimentally synthesized with a quenched two-dimensional quantum anomalous Hall system on a square Raman lattice. The quench dynamics for the quasimomentum-time-dependent Bloch vectors defines a Hopf map from $(q_x,q_y,t)\in T^3$ to the Bloch sphere $S^2$. In this Hopf map, a dynamical Hopf number can be defined, and it exactly equals the Chern number of the post-quench Hamiltonian.
We experimentally measure the Hopf link between the fibers for the North and South Poles on $S^2$, which are the trajectories in $(q_x,q_y,t)$ space with maximal spin polarization, to extract the topological Chern number of the post-quench Hamiltonian. We also observe the structure of Hopf fibration for the mutually nested Hopf tori. Our study sheds some new light on the interplay between topology (Hopf number) and geometry (fiber bundle) in quantum dynamics.
\end{abstract}

\maketitle

\begin{figure*}[htbp]
\begin{center}
\includegraphics[width=0.8\linewidth]{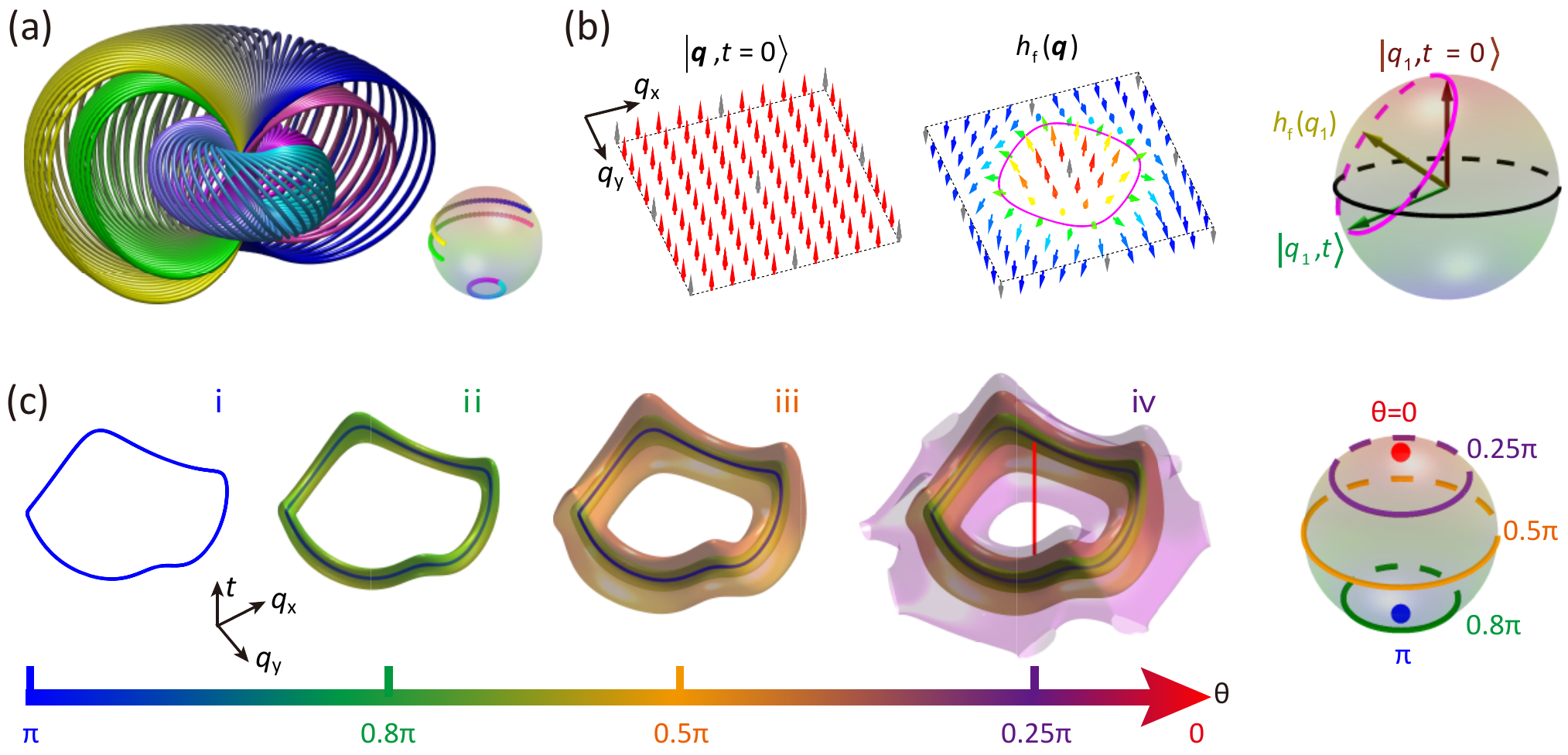}
\caption{
Illustration of the Hopf links and Hopf fibration.
(a) The Hopf fibration for the Hopf map from $S^3$ (left) to $S^2$ (right). 
(b) The quench dynamics for the QAH model, which features a Hopf map from $T^3$ to $S^2$~\cite{linking_HuiZhai}. We start with a spin polarized state $\left|\bm{q},t=0\right\rangle$ (left), then evolve the state by the QAH Hamiltonian $\bm{h}_{f}(\bm{q})\cdot\bm{\sigma}$ (middle). For each quasimomentum point $\bm{q}_1$, the time-dependent state vector precesses around the corresponding Hamiltonian vector $\bm{h}_{f}(\bm{q}_1)$ (right).
(c) For the Hopf map of the quenched QAH model, the fiber for the South (North) Pole on $S^2$ is a loop (straight line) in $T^3$. The fibers for each latitudinal circle on $S^2$ form a closed surface, which is homeomorphic to a torus; we thus term the closed surface as a Hopf torus. As we go from the South Pole to the north on $S^2$ by decreasing $\theta$, the corresponding Hopf tori in $T^3$ enclose each other one by one. The grow process for the Hopf tori nesting structure is shown in (i-iv).
}
\label{fig1}
\end{center}
\end{figure*}

Topological quantum matter has attracted great attention due to its intriguing physical properties and potential application in quantum computation and spintronics~\cite{topological_insulator,topo_insu_super,superconduct}. The topological classification of gapped free fermions for all dimensions based on the presence or absence of time-reversal, particle-hole, and chiral symmetries is also achieved with totally ten classes identified~\cite{Chiu2016}.
The Hopf insulator in three dimensions (3D)~\cite{Moore-Wen2008}, whose topology is determined by the Hopf number~\cite{Hopf1931},
apparently eludes the aforementioned classification scheme~\footnote{It can also be protected by certain symmetries other than time-reversal, particle-hole and chiral symmetries; see Ref.~\cite{Liu-Xu2017}.}.
Apart from topologically stable edge states~\cite{Moore-Wen2008,Duan2013}, a nonzero Hopf number also gives rise to the non-trivial Hopf links and Hopf fibration~\cite{Hopf1931} (see Fig.\ref{fig1}{(a)} as a schematic illustration of the Hopf fibration for the Hopf map from 3-sphere $S^3$ to 2-sphere $S^2$~\cite{quenchHaldane,Hopflinks,supplymentary}). It turns out that, the Hopf links has a close relationship with the Chern-Simons theory~\cite{Jones1985,Witten1989} and quantum gravity~\cite{Witten1988}. Moreover, the Hopf fibration is identical to the fiber bundle structure of the Dirac monopole~\cite{Dirac1931,Trautman1977}.

Recently, many well-known 2-dimensional (2D) topological quantum matters, such as the Haldane model on a hexagonal lattice~\cite{Haldane1988} and the quantum anomalous Hall model (QAH) on a square lattice~\cite{Qi2006}, have been synthesized with ultracold quantum gases using the shaking-lattice or Raman-lattice techniques~\cite{Esslinger_Haldane_model,sengstock_Berry_curvature,realization2DSOC}.
Nontrivial topological phases are also observed in these systems~\cite{W.S_longLive, uncover_topology,map2DSOCband}.
To date, although there exists certain proposal for realizing the Hopf insulator in a 3D optical lattice with ultracold atoms~\cite{Duan2018}, the corresponding experimental realization remains to be elusive. The study of the quench dynamics of a 2D two-band topological model (e.g., the Haldane model or the QAH model), however, sheds some new light~\cite{linking_HuiZhai}.
It is shown that, in this (2+1)-dimensional (i.e., 2D quasimomentum plus 1D time) space, a \emph{dynamical} Hopf number can be defined~\cite{linking_HuiZhai} and the \emph{dynamical} Hopf torus can be synthesized via Floquet engineering~\cite{2019arXiv190403202U}. 
Moreover, 
it is proved that the Chern number of the post-quench Hamiltonian can be achieved from the dynamical Hopf number, i.e., the linking number~\cite{linking_HuiZhai} and the winding number~\cite{2019arXiv190403202U}.
To a certain extent, this (2+1)-dimensional system can be regarded as a dynamical realization of a Hopf insulator, i.e, a \emph{dynamical Hopf insulator}. As a consequence of nonzero Hopf number, non-trivial Hopf links as well as Hopf fibration exist in this system, which can be observed with quasimomentum- and time-resolved time-of-flight image method~\cite{BlochReview2008}.
The Hopf link between the inverse images for the North Pole and the South Pole on the Bloch sphere $S^2$ has been measured by the Hamburg group in a Haldane-like model on a shaken hexagonal optical lattice~\cite{link2019}.

In this work, we use the quench dynamics of a 2D topological Raman lattice~\cite{Realization2DSOCtheory,W.S_longLive,realization2DSOC}, which in the tight-binding limit is mapped to the QAH model, to realize the dynamical Hopf insulator~\cite{Hopflinks}. We observe the Hopf link between the fibers for the North and South Poles on the Bloch sphere by measuring their trajectories $\mathcal{L_{+}}$ and $\mathcal{L_{-}}$ 
in $(q_x,q_y,t)$ space with maximal spin polarizations $P_z = \pm1$, and extract the Hopf number to determine the Chern number of the post-quench Hamiltonian.
The Hopf torus, which are the fibers for each latitudinal circle on the Bloch sphere, is also measured. The nesting structure of the Hopf tori for different latitudinal circles as the key information of the Hopf fibration, i.e., the topologically nontrivial way that the Hopf tori fill the whole (2+1)-dimensional space, is observed experimentally.

\emph{The Hopf map for the quenched QAH model}.--- Our system is based on a QAH model realized via a 2D spin-orbit (SO) coupled gas of ultracold ${}^{87}\textrm{Rb}$ atoms on a square lattice~\cite{W.S_longLive}. The 2D SO coupled system is described by the Hamiltonian
\begin{equation}
 H=\frac{\bm{p}^{2}}{2m}+V_{\rm{latt}}(x,y)+\frac{\delta}{2}\sigma_z+\Omega_{\rm{1}}(x,y)\sigma_x+\Omega_{\rm{2}}(x,y)\sigma_y,
 \label{realHamiltonian}
\end{equation}
where $\bm{p}$ and $m$ are the momentum and mass of the atom, respectively. $V_{\rm{latt}}(x,y)=V_{\rm{0}}(\cos^{2}k_{\rm{0}}x+\cos^{2}k_{\rm{0}}y)$ is the potential of the 2D optical lattice, where $V_{\rm{0}}$ is the lattice depth, and $k_0$ is the recoil momentum. $\delta$ is the two photon detuning. $\Omega_{\rm{1}}(x,y)=\Omega_{\rm{0}}{\cos}(k_{\rm{0}}x){\sin}(k_{\rm{0}}y)$ and $\Omega_{\rm{2}}(x,y)=\Omega_{\rm{0}}{\cos}(k_{\rm{0}}y){\sin}(k_{\rm{0}}x)$ are the Raman lattices, where $\Omega_{\rm{0}}$ is the Raman coupling strength.  $\bm{\sigma}=(\sigma_x,\sigma_y,\sigma_z)$ is the vector of Pauli matrices.

The two-band tight-binding model for Eq.~(\ref{realHamiltonian}) can be written as
$\mathcal{H}(\bm{q})=\bm{h}(\bm{q})\cdot\bm{\sigma}$,
where $\bm{h}(\bm{q})=(h_x, h_y, h_z)$ is a vector with $h_x=2t_{\rm{so}}\sin{q_y}$, $h_y=2t_{\rm{so}}\sin{q_x}$, $h_z=-\delta/2+2t_0\cos(q_x)+2t_0\cos(q_y)$, and $\bm{q}=(q_x,q_y)$ is the  quasimomentum~\cite{W.S_longLive, Realization2DSOCtheory}. Here, $t_{\rm{0}}$ and $t_{\rm{so}}$ are the spin-conserved and spin-flip tunneling coefficients, respectively.
We prepare the initial state as a spin-polarized state $\left|\psi(\bm{q},t=0)\right\rangle=\left|\uparrow\right\rangle$, and evolve the state by the Hamiltonian $\bm{h}_f(\bm{q})\cdot\bm{\sigma}$; see Fig.~\ref{fig1}(b). This quench process gives rise to a time-dependent state $\left| {\psi ({\bm{q}},t)} \right\rangle  = \exp ( - i{\bm{h}_f} \cdot \bm{\sigma} t)\left| {\psi ({\bm{q}},t = 0)} \right\rangle$; we take the reduced Planck constant $\hbar=1$ throughout. From this state, we can define a Bloch vector which lies on $S^2$ as
	${\mathbf{P}}({\bm{q}},t) = \left\langle {\psi ({\bm{q}},t)} \right|\bm{\sigma} \left| {\psi ({\bm{q}},t)} \right\rangle$.
The three components of ${\mathbf{P}}({\bm{q}},t)$ are given by
\begin{equation} \label{Eq:Hopf_map_T3_S2}
	\begin{aligned}
  {P_x}({\bm{q}},t) &= \sin ({t_{\bm{q}}}){{\hat h}_y} + [1 - \cos ({t_{\bm{q}}})]{{\hat h}_x}{{\hat h}_z}, \hfill \\
  {P_y}({\bm{q}},t) &=  - \sin ({t_{\bm{q}}}){{\hat h}_x} + [1 - \cos ({t_{\bm{q}}})]{{\hat h}_y}{{\hat h}_z}, \hfill \\
  {P_z}({\bm{q}},t) &= \cos ({t_{\bm{q}}}) + [1 - \cos ({t_{\bm{q}}})]\hat h_z^2, \hfill \\
\end{aligned}
\end{equation}
where $t_{\bm{q}}=2|\bm{h}_f(\bm{q})|t$ is the rescaled time, and ${\mathbf{\hat h}} = {(\hat h_x, \hat h_y, \hat h_z)} = {{\mathbf{h}}_f}/|{{\mathbf{h}}_f}|$ is the normalized post-quench Hamiltonian vector. There is a periodicity in the rescaled time direction, thus ${t_{\bm{q}}} \in [0,2\pi ) \cong {S^1}$. We then see that Eq.~(\ref{Eq:Hopf_map_T3_S2}) defines a Hopf map from $T^3=T^2\times S^1$ to $S^2$~\cite{linking_HuiZhai}. Here, the 2-torus $T^2$ stands for the first Brillouin zone (FBZ) of $(q_x, q_y)$.

\begin{figure*}[t]
\begin{center}
\includegraphics[width=0.85\linewidth]{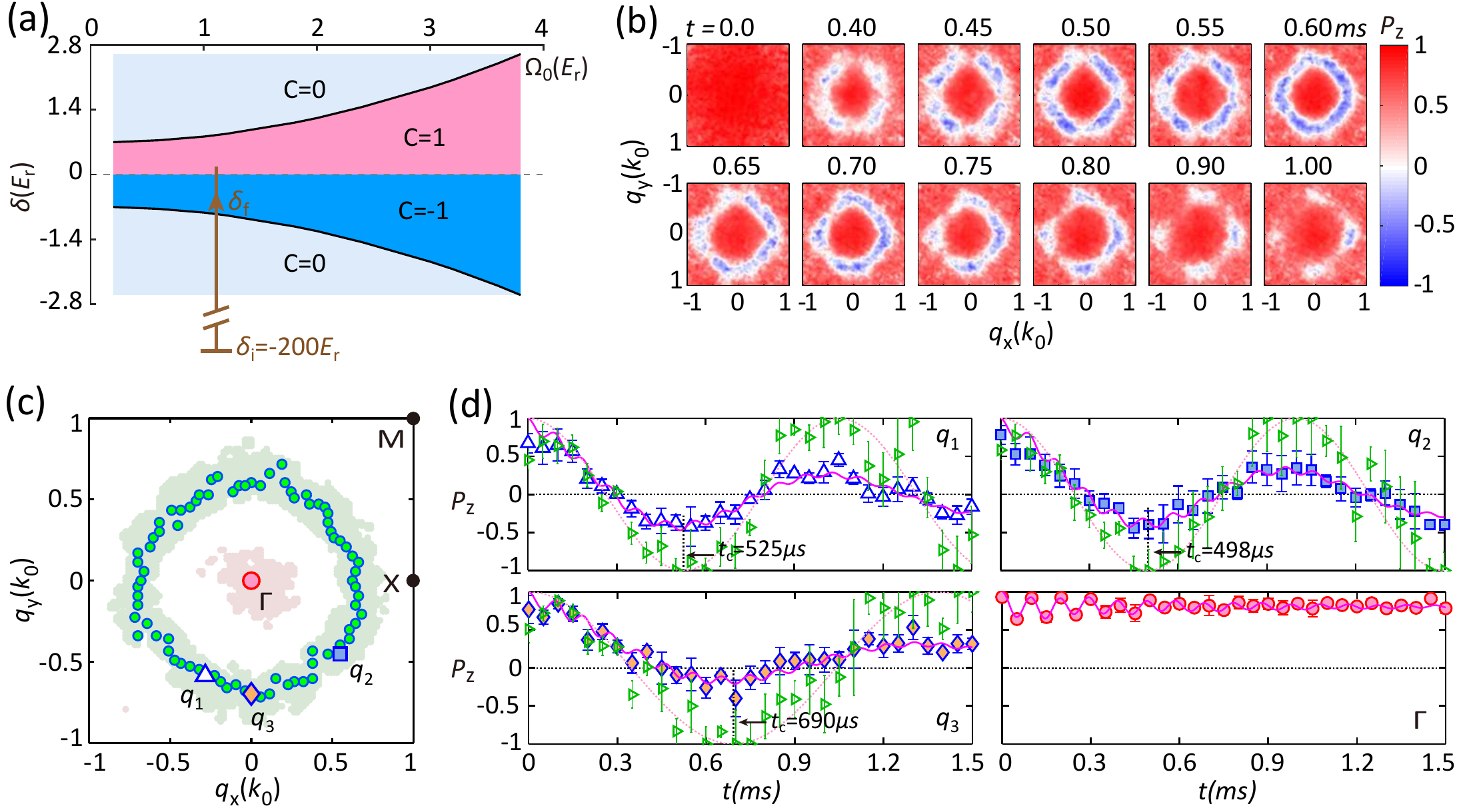}
\caption{
Quench dynamics for the spin polarization.
(a) The theoretical phase diagram (at $\delta$-$\Omega_{0}$ plane) of the lowest $s$-band.
The arrow indicates the quench from a trivial regime to a topological regime.  
(b) Time evolution of the experimentally measured spin polarization $P_{z}^{(\text{exp})}({\bm{q}},t)$ for $t=\SI{0}{\milli\second}$ to $\SI{1.0}{\milli\second}$. Here we take $V_{0}=4.0E_{{r}}, \Omega_{0}=1.0E_{{r}}$, and $\delta_{{f}}=0.2E_{{r}}$.
(c) The green circles are the $q$-points with largest spin-flip during the evolution. 
The green (red) shadow region shows the quasi-momentum points with spin polarizations below $-0.1$ (above $0.85$) during the evolution.
(d) The evolution of $P_{z}^{(\text{exp})}({\bm{q}},t)$ at selected $\bm{q}$, whose locations are marked in (c). The magenta solid curves are the fittings. The green triangles with error bars are experimental data after removing the damping and higher band effects~\cite{supplymentary}, and the dashed pink curves are the fitting with tight binding model. $t_{\rm{c}}$ is the time that $P_{z}^{(\text{exp})}({\bm{q}},t)$ reaches its first local minimum.
}
\label{fig2}
\end{center}
\end{figure*}

A Hopf number $I_H$ can be defined for this Hopf map~\cite{Hopflinks,supplymentary}; and it has been proved that, this Hopf number equals the Chern number of (the lowest band of) the post-quench Hamiltonian~\cite{linking_HuiZhai}. The Hopf fibration of this Hopf map is shown schematically in Fig.~\ref{fig1}(c). Instead of measuring the fiber in $T^3$ for each point on the Bloch sphere $S^2$, which generally requires a full Bloch-state tomography~\cite{Alba2011,Monroe2017}, here we propose to measure the collection of fibers (which is a tube that is homeomorphic to a torus in $T^3$, i.e., the Hopf torus) that map to a particular latitudinal circle on $S^2$. The parametric equation for the Hopf torus is given by:
\begin{equation}
 	{P_z}({\bm{q}},t) = \cos \theta ,\label{Hopf_map_z}
 \end{equation}
 where $\theta$ is the polar angle of the Bloch sphere. In the following, we demonstrate our experimental results to extract the fibers for ${P_z}({\bm{q}},t) = \pm1$ [the blue loop and red line in Fig.~\ref{fig1}(c)], and use the linking number between them to map out the topology of the post-quench Hamiltonian. We also experimentally measure the nesting Hopf tori structure in Fig.~\ref{fig1}(c).

\emph{Quench dynamics in the experiment}.---
In the experiment, the ${}^{87}\textrm{Rb}$ atoms are firstly pumped to a spin up state $\left|\uparrow\right\rangle=\left|F=1,m_F=-1\right\rangle$ and cooled above the critical temperature of the Bose-Einstein condensate. Subsequently, the atoms are loaded into the 2D SO coupling Hamiltonian [Eq.~(\ref{realHamiltonian})] adiabatically with initial detuning $\delta_{{i}}=-200E_{{r}}$ [here $E_r=k_0^2/(2m)$ is the recoil energy] and reach the thermal equilibrium. The Raman couplings $\Omega_1$ and $\Omega_2$ are effectively suppressed due to the large detuning, and the atoms feel \emph{only} the 2D lattice potential $V_{\rm{latt}}$. The quench is executed by switching the detuning from $\delta_{{i}}$ to the final near-resonant value $\delta_{{f}}$ within $\SI{200}{\nano\second}$, as indicated by the arrow in Fig.~\ref{fig2}(a).
The quench process activates a non-equilibrium evolution of Raman-induced spin oscillations
between $\left|\uparrow\right\rangle$ and $\left|\downarrow\right\rangle$ (=$\left|1,0\right\rangle$) states governed by $H(\delta_f)$. The spin oscillation is quantified by the time-evolution of the quasimomentum dependent spin-polarization $P_{{z}}^{(\text{exp})}({\bm{q}},t)=\frac{N_{\uparrow}-N_{\downarrow}}{N_{\uparrow}+N_{\downarrow}}$, where $N_{\uparrow}$ ($N_{\downarrow}$) denotes the number of atoms at spin up (down) state in the quasimomentum space (see~\cite{uncover_topology} for details).

The dynamical evolution of $P_{{z}}^{(\text{exp})}({\bm{q}},t)$ in the FBZ is shown in Fig.~\ref{fig2}(b) for a typical experiment.
It shows a {\lq\lq}ring" structure when $H(\delta_{{f}})$ locates in the topological region~\cite{uncover_topology}. On the ring, $P_{{z}}^{(\text{exp})}({\bm{q}},t)$ can flip from positive to negative.
Away from the ring, especially at the high symmetric points ($\Gamma, M$ and $X$), $P_{{z}}^{(\text{exp})}({\bm{q}},t)$ keeps positive during the period of unitary evolution.
The quasimomentum points with maximal spin-flip are presented on the ring in Fig.~\ref{fig2}(c).
The time evolution of the selected quasimomenta $\bm{q}_{1,2,3}$ and $\Gamma$ are shown in Fig.~\ref{fig2}(d).

\begin{figure*}[t]
\begin{center}
\includegraphics[width=0.8\linewidth]{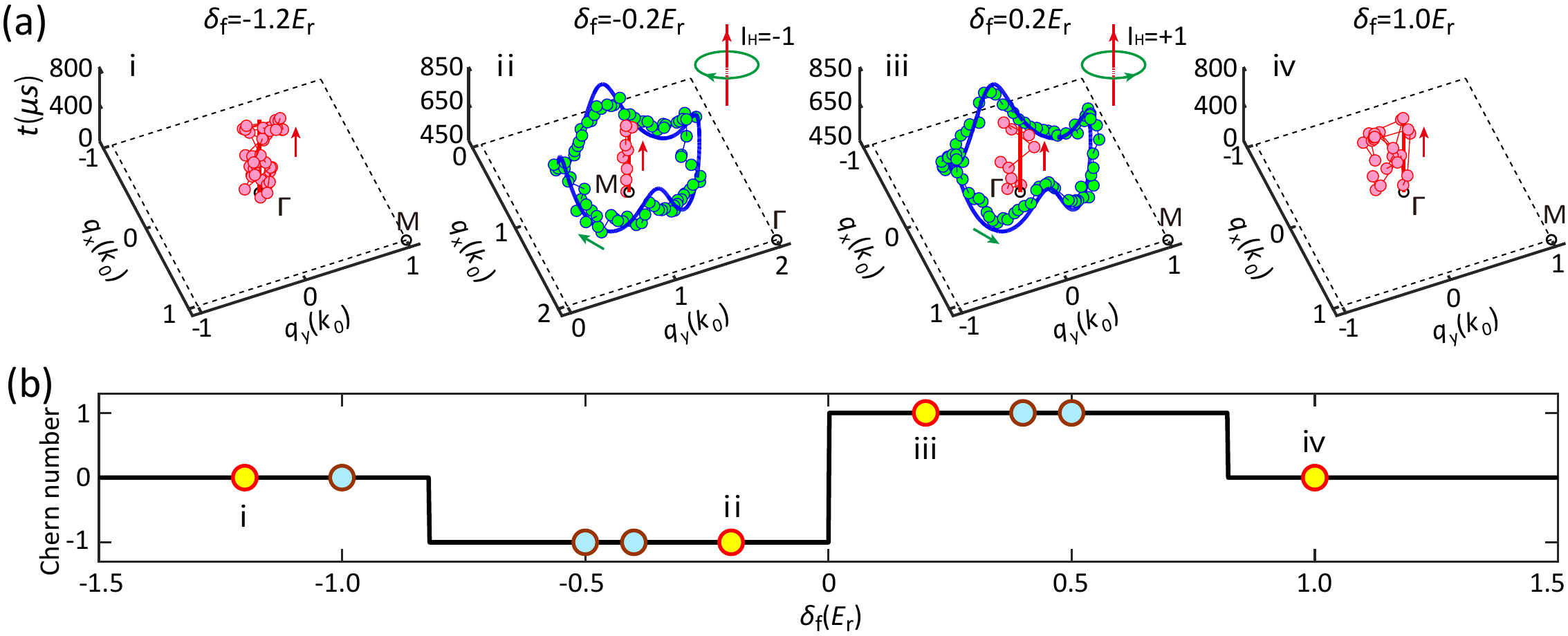}
\caption{
Determining the Chern number from the Hopf links of $P_z = \pm1$ in $(\bm{q},t)$ space.
(a) The dynamical trajectories $\mathcal{L_+}$ for $P_z=+1$ (red dots) and $\mathcal{L_-}$ for $P_z=-1$ (blue dots) in $(\boldsymbol{q},t)$ space. The subfigures from {\romannumeral1} to {\romannumeral4} show the trajectories with different final detunings $\delta_f$ for the case of $\Omega_{\rm{0}}=1.0E_{\rm{r}}$.
The solid lines are from the theoretical calculation and the dots are from the experiments. 
The sign of the Hopf linking number~\cite{supplymentary} can be read out using the right-hand rule, as shown schematically in the top right corners of ii and iii.
(b) By varying the final detuning $\delta_f$, the Chern number of the lowest band also changes, which exactly matches the change of the Hopf number as extracted from (a).
}
\label{fig3}
\end{center}
\end{figure*}

\emph{Obtaining the trajectories $\mathcal{L_{\pm}}$ and the Hopf linking number in quasimomentum-time space}.---
In order to get the trajectories $\mathcal{L_+}$/$\mathcal{L_-}$ of spin up/down states (i.e., the North/South Poles on the Bloch sphere), the quasimomentum points for which the spin polarization can reach $\pm1$ during the dynamical evolution and the time that $P_{{z}}({\bm{q}},t)=\pm1$ are recorded.
The dynamical evolution of the spin polarizations at $\bm{q}_{1,2,3}$ is found to be a damped oscillation with multi-frequency components as shown in Fig.~\ref{fig2}(d). It is caused by the dephasing-induced damping and multi-band effect in the real system~\cite{uncover_topology}.
To map the results back to the two-band tight-binding model, we have thus extracted and removed the dephasing and high frequency components~\cite{supplymentary}. And the unitary evolutions of spin polarization between two lowest bands are revealed, as shown by the normalized triangle data in Fig.~\ref{fig2}(d).
From the fitting, the time $t_{{c}}(\bm{q}_{1,2,3})$ of $P_{{z}}$ reaches $-1$ are recorded as $525\mu s$, $498\mu s$, and $690\mu s$, respectively.

By applying the above recipe, all the quasimomenta with $P_{{z}}(\bm{q},t)$ reaching $-1$ are obtained on the ring, presented as dark green circles in Fig.~\ref{fig2}(c), together with their time $t_{{c}}$.
By plotting these quasimomenta and the corresponding time $t_{{c}}$ in the 2-dimensional quasimomentum plus 1-dimensional time space $(q_x, q_y, t)$, the trajectory $\mathcal{L_-}$ is achieved, as shown by the connected green dots which form a closed loop in Fig.~\ref{fig3}(a).{\romannumeral3}.
Close to the $\Gamma$ point, $P_{{z}}(\bm{q},t)$ between two lowest band exhibits constant value with $P_{{z}}=+1$ during the dynamical evolution. So the trajectory $\mathcal{L_+}$ is almost a straight line in the $(q_x, q_y, t)$ space, as shown by the connected red dots in Fig.~\ref{fig3}(a).{\romannumeral3}.

From Fig.~\ref{fig3}(a).{\romannumeral3}, we see that the trajectory of $\mathcal{L_-}$ winds around $\mathcal{L_+}$. $\mathcal{L_+}$ behaves as a straight line, which is actually a circle in $T^3$ because of periodic boundary condition along the time direction. Thus we see  that $\mathcal{L_+}$ also circles around $\mathcal{L_-}$ simultaneously. This structure forms a link called the Hopf link, with Hopf (linking) number $I_{{H}}=+1$~\cite{linking_HuiZhai}.
By taking $V_{0}=4.0E_{{r}}, \Omega_{0}=1.0E_{{r}}$, and varying the final detuning $\delta_{{f}}$, we can get the Hopf links and linking number in different conditions as presented in Fig.~\ref{fig3}(a).
For $\delta_{{f}}=-0.2E_{{r}}$ and $+0.2E_{{r}}$, the Hopf linking number is $-1$ and $+1$, respectively. It corresponds to the Chern number ${{C}}=-1$ and $+1$ for the 2D QAH system, as shown in Fig.~\ref{fig3}(b), which has topological band structures~\cite{realization2DSOC,W.S_longLive,uncover_topology,supplymentary}.
For $\delta_{{f}}=-1.2$ and $+1.0E_{{r}}$, only the trajectories $\mathcal{L_+}$ can be extracted and the Hopf linking number vanishes, corresponding to trivial band structure in the 2D QAH system with ${C}=0$ (see \cite{supplymentary} for more details).

To further demonstrate validity of the method, the two trajectories $\mathcal{L_{\pm}}$ calculated by Eq.~\ref{Hopf_map_z} with $\theta=0$ and $\pi$ are drawn as the solid curves in Fig.~\ref{fig3}(a).{\romannumeral1}-{\romannumeral4}. The calculated results agree well with the experiment measurements.
The change of the Chern number as the final detuning $\delta_f$ are shown in Fig.~\ref{fig3}(b), which exactly matches the change of the Hopf linking number as extracted from Fig.~\ref{fig3}(a). The topology of the post-quench Hamiltonian is identified correspondingly.

\emph{Observation of nested Hopf tori}.---
In addition to getting the trajectories of $\mathcal{L_\pm}$ and the Hopf linking number between them,
the dynamical evolution of the spin polarization in the FBZ after quench also contains the information to observe the nested Hopf tori.
As shown in Fig.~\ref{fig1}(c), each latitudinal circle on the Bloch sphere is mapped as a closed surface in the quasimomentum-time space.
According to the Hopf map in Eq.~(\ref{Eq:Hopf_map_T3_S2}), a given latitudinal circle with polar angle $\theta$ corresponds to a series of quasimomentum points $\bm{q}$ in the FBZ, whose time-dependent spin polarization can reach $P_z(\bm{q},t)=\cos{\theta}$ during the evolution at certain times.
The quasimomenta $\bm{q}$ and the time $t_c$ are extracted using the same approach as the one to extracting $\mathcal{L_{\pm}}$ (see \cite{supplymentary} for details).

\begin{figure}[htbp]
\begin{center}
\includegraphics[width=\linewidth]{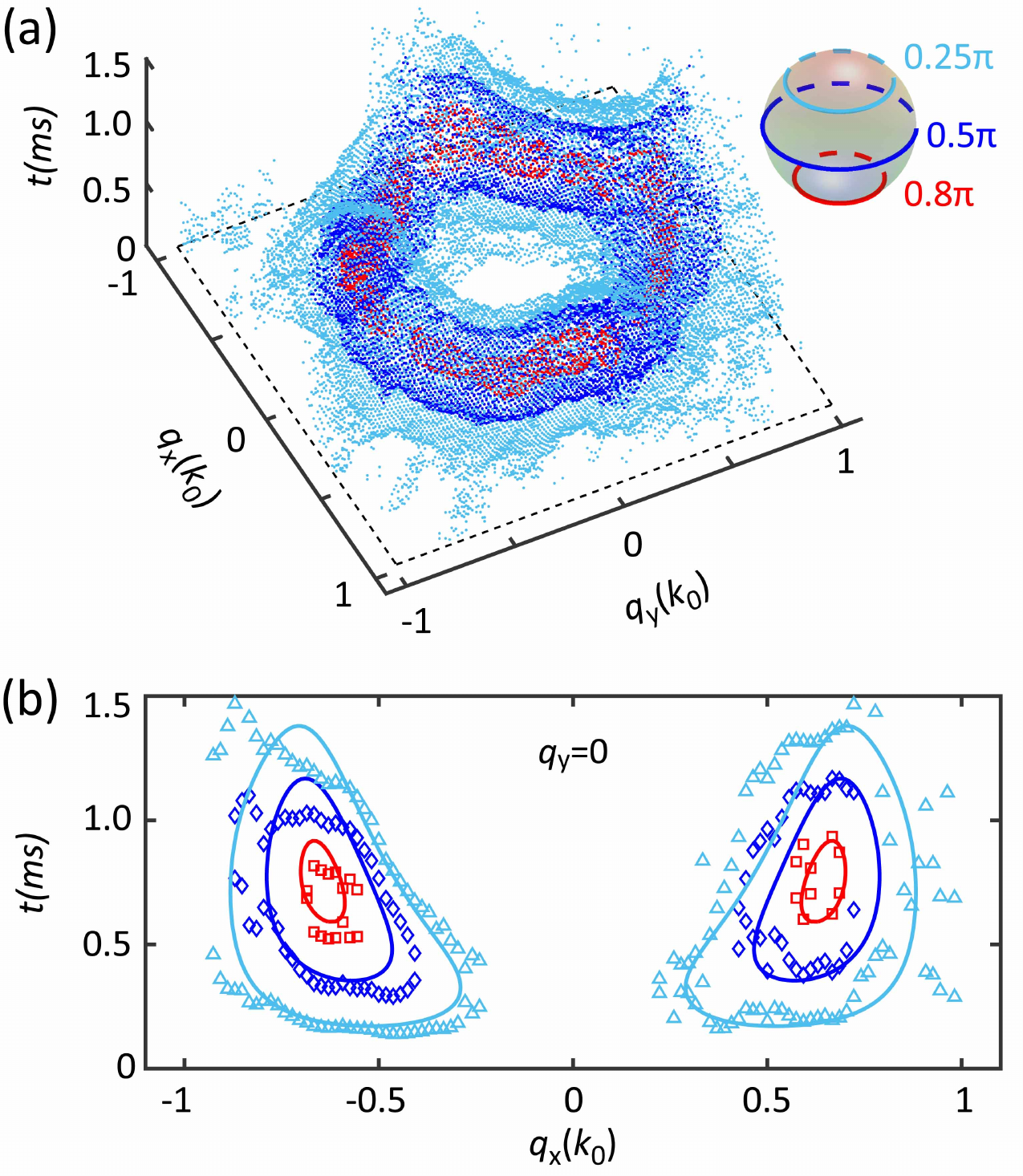}
\caption{
Observation of the nested Hopf tori.
(a) Hopf tori as the inverse images of the latitudinal circles on the Bloch sphere (top right corner) with $\theta=0.8\pi$ (red), $0.5\pi$ (blue), and $0.25\pi$ (cyan), respectively. The torus for larger polar angle $\theta$ is completely enclosed by the tori for smaller polar angles. Here we take $V_{{0}}=4.0E_{r}, \Omega_{0}=1.0E_{r}, \delta_{f}=0.2E_{{r}}$ (same as in Fig.~\ref{fig3}{(a)}.iii).
(b) Sectional view of (a) along $q_{y}=0$, from which the nesting structure of the Hopf tori can be identified more transparently. The dots are the experimental data, the lines are the theoretical results.}
\label{fig4}
\end{center}
\end{figure}

Figure \ref{fig4} shows the construction of the Hopf tori in our experiment with parameters $V_{0}=4.0E_{{r}}, \Omega_{0}=1.0E_{{r}}$ and $\delta_{{f}}=0.2E_{{r}}$. For the latitudinal circle with $\theta=0.25\pi$, the quasimomenta $\bm{q}$ with time $t_c$ that satisfy the formula $P_z(\bm{q},t_c)=\cos{\theta}$ are presented by cyan dots in Fig.~\ref{fig4}(a). They form a close surface with a hole, i.e., a torus.
For $\theta=0.5\pi$ and $0.8\pi$, the corresponding Hopf tori are shown in Fig.~\ref{fig4}{(a)} by blue and red dots, respectively.
It is clear to see that the torus for larger polar angle $\theta$ is completely enclosed by the tori for smaller polar angles.
Figure \ref{fig4}{(b)} presents the sectional view of the Hopf tori along the plane $q_y=0$. It is also clear that the tori with large polar angle are enclosed by that with smaller ones.
These features indicate that a dynamical Hopf insulator with a nontrivial Hopf number is constructed.

\emph{Conclusion and discussion}---
In summary, we have experimentally synthesized a 3D dynamical Hopf insulator using the quench dynamics of a 2D quantum anomalous Hall system with ultracold bosons loaded into a 2D Raman lattice.
In the space of 2D quasimomentum plus 1D time, the trajectories of spin up and down atoms and their linking number are achieved to extract the topology of the post-quench Hamiltonian.
Furthermore, by applying the Hopf map, the Hopf fibration structure of mutually nested Hopf tori, which characterizes the topological feature of the dynamical Hopf insulator, is constructed. 
Our experiment exhibits the true power of quantum simulation with ultracold quantum gases to reveal the geometric structure of topological quantum state of matter using quantum dynamics.

\emph{Acknowledgement}---
We thank Hui Zhai, Youjin Deng and Xiong-Jun Liu for fruitful discussions. This work was supported by the National Key R\&D Program of China (under grant No.~2016YFA0301601), National Nature Science Foundation of China (under grants No.~11674301), and the Anhui Initiative in Quantum Information Technologies (AHY120000).

%

\newpage
\onecolumngrid
\renewcommand\thefigure{S\arabic{figure}}
\setcounter{figure}{0}
\renewcommand\theequation{S\arabic{equation}}
\setcounter{equation}{0}
\newpage
{
\center \bf \large
Supplemental Material for: \\
Observation of the Hopf Links and Hopf Fibration in a 2D topological Raman Lattice\vspace*{0.1cm}\\
\vspace*{0.0cm}
}
\begin{center}
Chang-Rui Yi$^{1,2}$, Jin-Long Yu$^{3,4}$, Wei Sun$^{1,2}$, Xiao-Tian Xu$^{1,2}$, Shuai Chen$^{1,2,*}$ and Jian-Wei Pan$^{1,2,\dagger}$\\
\vspace*{0.15cm}
\small{\textit{$^1${Hefei National Laboratory for Physical Sciences at Microscale and Department of Modern Physics, University of Science and Technology of China, Hefei, Anhui 230026, China}\\
$^2$Shanghai Branch, CAS Center for Excellence and Synergetic Innovation Center in Quantum Information and Quantum Physics, University of Science and Technology of China, Shanghai 201315, China\\
$^3${Institute for Advanced Study, Tsinghua University, Beijing 100084, China}\\
$^4${Institute for Quantum Optics and Quantum Information of the Austrian Academy of Sciences, Innsbruck A-6020, Austria}\\
}}
\vspace*{0.25cm}
\end{center}

This supplementary material provides more details
for the various points mentioned in the main article.

\subsection{On the details of the Hopf map from $S^3$ to $S^2$ and its visualization in $\mathbb{R}^3$}

The map proposed by Heinz Hopf in Ref.~\cite{Hopf1931} is a map from 3-sphere $S^3$ to 2-sphere $S^2$. In contrast, the Hopf map that studied in this work as shown in Eq.(2) is a map from 3-torus $T^3$ to $S^2$. In the case that the map of any subtorus $T^2$ [i.e., $(q_x, q_y)\in T^2$, $(q_x, t)\in T^2$, and $(q_y, t)\in T^2$] to $S^2$ takes zero Chern number [i.e., zero flux of Berry curvature, as shown in Eq.~(\ref{Eq:Berry_curvature}) below, penetrating the corresponding perpendicular subtorus], as is the case studied here, the classification of the map from $T^3$ to $S^2$ can be reduced to the one of $S^3$ to $S^2$~\cite{Moore-Wen2008}.

In order to get some intuitive understanding of the Hopf fibration, in this section, we review some well known results of the Hopf map from $S^3$ to $S^2$ in a pedagogical manner; more details can be found in Ref.~\cite{Nakahara2003book}.

The unit n-sphere $S^n$ is embedded in the Euclidean space $\mathbb{R}^{n+1}$ as
\[{({x_1})^2} + {({x_2})^2} + ... + {({x_{n + 1}})^2} = 1,\]
where $x_i$, $i=1,2,...,(n+1)$, is the coordinate in $\mathbb{R}^{n+1}$. For the 3-sphere $S^3$ in $\mathbb{R}^4$, it takes the following form,
\[{({x_1})^2} + {({x_2})^2} + {({x_3})^2} + {({x_4})^2} = 1.\]
We can introduce two complex coordinates ${z_0} = {x_1} + i{x_2}$ and ${z_1} = {x_3} + i{x_4}$, and then it becomes
\[|{z_0}{|^2} + |{z_1}{|^2} = 1.\]
We parameterize $S^2$ as
\[{({\xi _1})^2} + {({\xi _2})^2} + {({\xi _3})^2} = 1.\]
The Hopf map $\pi :\;{S^3} \to {S^2}$ is defined as
\begin{equation} \label{Eq:Hopf_map_S3_S2}
	\begin{gathered}
  {\xi _1} = 2({x_1}{x_3} + {x_2}{x_4}), \hfill \\
  {\xi _2} = 2({x_2}{x_3} - {x_1}{x_4}), \hfill \\
  {\xi _3} = {({x_1})^2} + {({x_2})^2} - {({x_3})^2} - {({x_4})^2}. \hfill \\
\end{gathered}
\end{equation}
It can be verified that $\pi$ maps $S^3$ to $S^2$ because
\[{({\xi _1})^2} + {({\xi _2})^2} + {({\xi _3})^2} = {[{({x_1})^2} + {({x_2})^2} + {({x_3})^2} + {({x_4})^2}]^2} = 1.\]
Let $(X,Y) = (\frac{{{\xi _1}}}{{1 - {\xi _3}}},\frac{{{\xi _2}}}{{1 - {\xi _3}}})$ be the stereographic coordinates in $\mathbb{R}^2$ of a point on $S^2$ projected from the North Pole.
Then the Hopf map in Eq.~(\ref{Eq:Hopf_map_S3_S2}) can be reformulated by the complex coordinates $Z=X+i Y$, $z_0= {x_1} + i{x_2}$, and $z_1= {x_3} + i{x_4}$ as follows:
\begin{equation}
	Z = \frac{{{\xi _1} + i{\xi _2}}}{{1 - {\xi _3}}} = \frac{{{x_1} + i{x_2}}}{{{x_3} + i{x_4}}} = \frac{{{z_0}}}{{{z_1}}}.
\end{equation}
We see that $Z$ is invariant under
\[({z_0},{z_1}) \mapsto ({e^{i\chi }}{z_0},{e^{i\chi }}{z_1}),\]
where $\chi\in[0,2\pi)$ is a real parameter. For a given value $Z^* = z^*_0 / z^*_1$ that represents a particular point on $S^2$ that is mapped from another point in $S^3$ with parameters $(z^*_0, z^*_1)$,  the points $({e^{i\chi }}{z^*_0},{e^{i\chi }}{z^*_1})$ in $S^3$ all map to the same point in $S^2$ that corresponds to $Z^* $: they form the fiber in $S^3$, and the fiber is a circle that is parametrized by the angle variable $\chi\in[0,2\pi)$.

We can directly parameterize $S^3$ as follows:
\[\begin{aligned}
  {z_0} &= {e^{i\chi }}{e^{i\phi /2}}\cos (\theta /2), \hfill \\
  {z_1} &= {e^{i\chi }}{e^{ - i\phi /2}}\sin (\theta /2), \hfill \\
\end{aligned} \]
where the angle variables $\chi$ and $\phi$ take values between $0$ and $2\pi$, and $\theta$ takes value between $0$ and $\pi$. And the corresponding coordinates in $\mathbb{R}^4$ is given by
\begin{equation}\label{Eq:Hopf_fibers_R4}
	\begin{aligned}
  {x_1} &= \cos (\chi  + \frac{\phi }{2})\cos (\theta /2),\quad &{x_2}& = \sin (\chi  + \frac{\phi }{2})\cos (\theta /2), \hfill \\
  {x_3} &= \cos (\chi  - \frac{\phi }{2})\sin (\theta /2),\quad &{x_4} &= \sin (\chi  - \frac{\phi }{2})\sin (\theta /2). \hfill \\
\end{aligned}
\end{equation}
Then the Hopf map $Z=z_0/z_1$ gives rise to
\[Z = \frac{{{z_0}}}{{{z_1}}} = {e^{i\phi }}\cot (\theta /2) = \frac{{\sin \theta }}{{1 - \cos \theta }}\cos \phi  + i\frac{{\sin \theta }}{{1 - \cos \theta }}\sin \phi ,\]
from which the coordinates that parameterize $S^2$ in $\mathbb{R}^3$ can be identified as
\[{\xi _1} = \sin \theta \cos \phi ,\;{\xi _2} = \sin \theta \sin \phi ,\;{\xi _3} = \cos \theta .\]
This expression is just the conventional parametrization for the 2-sphere $S^2$. For a given point on $S^2$ with angle variables $(\theta, \phi)$, the corresponding fiber in $S^3$ is parameterized by angle variables  $(\theta, \phi, \chi)$ according to Eq.~(\ref{Eq:Hopf_fibers_R4}), with $\chi$ running from $0$ to $2\pi$. Equation (\ref{Eq:Hopf_fibers_R4}) shows a point on $S^3$ embedded in $\mathbb{R}^4$, which could not be directly visualized in $\mathbb{R}^3$. To solve this problem, we again adopt the stereographic coordinates $(\frac{{{x_1}}}{{1 - {x_4}}},\frac{{{x_2}}}{{1 - {x_4}}},\frac{{{x_3}}}{{1 - {x_4}}})$ in $\mathbb{R}^3$ for a point on $S^3$, and the results are represented in Fig.1\textbf{a} in the main text.

\subsection{On the determination of the fiber's direction to get the sign of the linking number}
As stated in the main text, for the quench dynamics starting with a topologically trivial state, e.g., the spin polarized state $\left|\Psi(\bm{q},t=0)\right\rangle=\left|\uparrow\right\rangle$, the evolution by the post-quench Hamiltonian $\bm{h}_f(\bm{q})\cdot\bm{\sigma}$ gives rise to the quasimomentum-time-dependent state as $\left| {\psi ({\bm{q}},t)} \right\rangle  = \exp ( - i{\bm{h}_f} \cdot \bm{\sigma} t)\left| {\psi ({\bm{q}},t = 0)} \right\rangle$. For such a state, we can define the Berry connection $A_\mu$ and Berry curvature $J_\mu$ as follows~\cite{Wilczek_Zee1983,linking_HuiZhai}:
\begin{equation}
	\begin{aligned}
  {A_\mu }({\bm{q}},t) &= \frac{i}{{2\pi }}\left\langle {\psi ({\bm{q}},t)} \right|{\partial _\mu }\left| {\psi ({\bm{q}},t)} \right\rangle , \hfill \\
  {J_\mu } &= {\epsilon _{\mu \nu \lambda }}{\partial _\nu }{A_\lambda }, \hfill \\
\end{aligned}
\end{equation}
where the index $\mu$ takes value in $q_x$, $q_y$, and $t$, and ${\epsilon _{\mu \nu \lambda }}$ is the totally antisymmetric tensor. It can be shown that, the Berry curvature $J_\mu$ can be expressed as a local function of the Bloch vector ${\mathbf{P}}({\bm{q}},t) = \left\langle {\psi ({\bm{q}},t)} \right|\bm{\sigma} \left| {\psi ({\bm{q}},t)} \right\rangle$ as follows~\cite{Wilczek_Zee1983}:
\begin{equation}\label{Eq:Berry_curvature}
	{J_\mu } = \frac{1}{{8\pi }}{\epsilon _{\mu \nu \lambda }}{\mathbf{P}} \cdot ({\partial _\nu }{\mathbf{P}} \times {\partial _\lambda }{\mathbf{P}}).
\end{equation}
The vector of the Berry curvature $(J_{q_x}, J_{q_y}, J_t)$ defined in the $(\bm{q},t)$ space is the tangent vector of the fiber for the Bloch vector ${\mathbf{P}}({\bm{q}},t)$ on $S^2$, which gives a consistent definition of the direction of the fiber~\cite{linking_HuiZhai}. In the following, to show explicitly the periodicity in the time direction, we rescale the time direction by introducing the rescaled time coordinate ${\tilde t}({\bm{q}})=t_{\bm{q}}=2|\bm{h}_f(\bm{q})|t$. The dynamical Hopf number $I_H$ can be defined as follows
\begin{equation}
	{I_H} = \int_{ - \pi }^\pi  {d{q_x}} \int_{ - \pi }^\pi  {d{q_y}} \int_0^{2\pi } {d\tilde t} \sum\limits_{\mu  = {q_x},{q_y},\tilde t} {{A_\mu }{J_\mu }} .
\end{equation}
The geometric meaning of the Hopf number is the linking number between the fibers in $T^3$ for any two points on $S^2$. We may identify these two quantities, and sometimes term them as a whole -- the Hopf linking number.

In the $(\bm{q},{\tilde t})$ space, for the $\Gamma$ and $M$ points in the FBZ, we have $J_{q_x}=0$, $J_{q_y}=0$, but
\begin{equation}
	\begin{gathered}
  {J_{\tilde t}}({q_x} = 0,{q_y} = 0,\tilde t) = \frac{{16{{\sin }^2}(\tilde t/2)t_{{\text{so}}}^2}}{{{{(\delta  - 8{t_0})}^2}\pi }} \geqslant 0, \hfill \\
  {J_{\tilde t}}({q_x} = \pi ,{q_y} = \pi ,\tilde t) = \frac{{16{{\sin }^2}(\tilde t/2)t_{{\text{so}}}^2}}{{{{(\delta  + 8{t_0})}^2}\pi }} \geqslant 0. \hfill \\
\end{gathered}
\end{equation}

\begin{figure}[htbp]
\begin{center}
\includegraphics[width=0.75\linewidth]{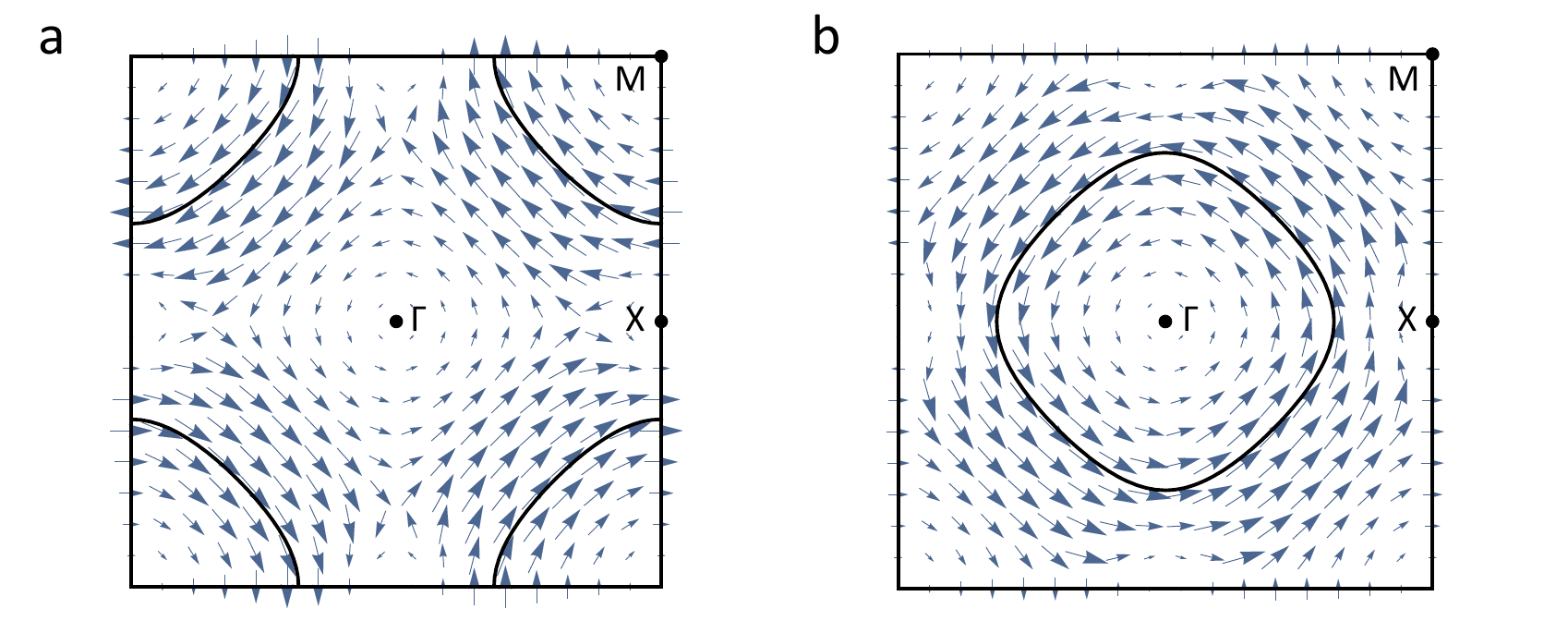}
\caption{
{\textbf{Determination of the direction of the fiber for the South Pole.} Here we show the distribution of the vector $({J_q}_x, {J_q}_y)$ in the $(q_x, q_y, {\tilde t})$ space at rescaled time ${\tilde t}=\pi$ for the Chern number $C=-1$ case (\textbf{a}), and $C=1$ case (\textbf{b})}. In particular, we set $t_\text{so}=t_0$ here, and take $\delta=-2.4t_0$ in \textbf{a}, and $\delta=2.4t_0$ in \textbf{b}. The contours in \textbf{a} and \textbf{b} are determined by the equation $h_z=-\delta/2+2t_0\cos(q_x)+2t_0\cos(q_y)=0$, which are just the fibers for the South Pole. To be more accurate, the fibers are the corresponding contours lie on the ${\tilde t}=\pi$ plane: the Bloch vectors on the contours all point to the South Pole direction at ${\tilde t}=\pi$. We can see that, the fiber winds around the $M$ point along the clockwise direction for the $C=-1$ case (\textbf{a}); it winds around the $\Gamma$ point along the counterclockwise direction for the $C=1$ case (\textbf{b}). The directions of the fibers for the South Pole shown here are also indicated by the green arrows in Fig.3{\textbf{a}}.ii and iii in the main text.
}
\label{figS1_t}
\end{center}
\end{figure}

They are all nonnegative functions, which indicate that the directions of the fibers for the North Pole that located at $\Gamma$ and $M$ all point to the positive time direction, as shown by the red arrows in Fig.3\textbf{a}. To further determine the direction of the fiber for the South Pole, we first note that, through direct evaluation, we have ${J_{\tilde t}}({q_x} = q_x^*,{q_y} = q_y^*,\tilde t = \pi ) = 0$, where $(q_x^*, q_y^*)$ is the contour in the FBZ determined by the vanishing of $h_z$ component (as an example, see the red loop in the middle of Fig.1\textbf{b}). We then see that, the vector $\left({J_q}_x(q_x^*, q_y^*, {\tilde t}=\pi), {J_q}_y(q_x^*, q_y^*, {\tilde t}=\pi)\right)$ determines the direction of the fiber for the South Pole, as shown schematically in Fig.~\ref{figS1_t}.
For the case that the trajectories $\mathcal{L}_+$ and $\mathcal{L}_-$ are linked once, the Hopf linking number $I_H=\pm1$ when the direction of the fiber $\mathcal{L}_+$ is the same/opposite as the normal direction of the fiber $\mathcal{L}_-$ determined by the right-hand rule. Thus we see that, for the $\delta_f=-0.2E_r$ case, $I_H=-1$ (Fig.3\textbf{a}.{\romannumeral2}); for the $\delta_f=+0.2E_r$ case, $I_H=+1$ (Fig.3\textbf{a}.{\romannumeral3}).

\subsection{On the quench dynamics of spin polarization in two bands model and real system}
\begin{figure}[htbp]
\begin{center}
\includegraphics[width=0.5\linewidth]{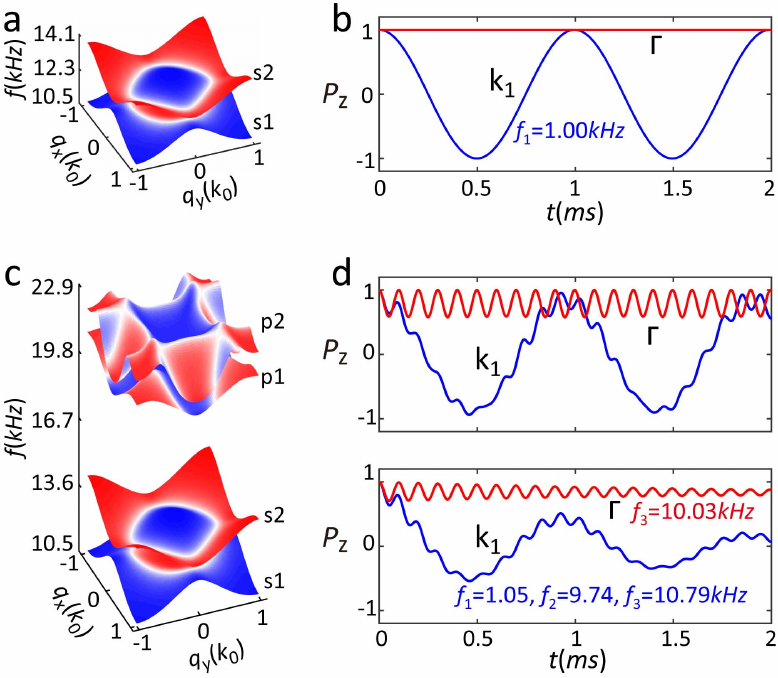}
\caption{
\textbf{The quench dynamics of spin polarization in two bands model and real system.}
\textbf{a}, The two-band structure including two $s$-band (s1,s2).
\textbf{b}, The dynamical evolution of $\Gamma$ and $k_1$ ($0.38\pi,-0.44\pi$) points.
\textbf{c}, The multi-band structure including two $s$-band (s1, s2) and two higher bands (p1,p2) in equilibrium state. The parameters are $V_{0}=4.0E_{{r}}, \Omega_{0}=1.0E_{{r}}$ and $\delta_{{f}}=0.2E_{{r}}$.
\textbf{d}, The dynamical evolution of $\Gamma$ and $k_1$ points calculated by numerical simulation in the real Hamiltonian.
}
\label{figS2}
\end{center}
\end{figure}

In the two-band tight-binding model, according to Eq.(2) in the main text, the time evolution of the spin polarization is a unitary evolution:
\begin{equation}
P_{{z}}(\bm{q},t)=A_1\cos(\omega_1(\bm{q}) t)+D(\bm{q}),\label{ideal_P}
\end{equation}
where $A_1=\hat h_{x}^{2}+\hat h_{y}^{2}$ is the amplitude, $D(\bm{q})=\hat h_{z}^{2}$ is the offset, and $\omega_1({\bm{q}})=2\pi f_1=2\left|h_f\right|$ is the angular frequency.
The oscillation frequency at a certain momentum point is $f_1={\left| h_f(\bm{q})\right|}/{\pi}$; the existence of this {\emph{single}} frequency, instead of multi frequencies, is due to the two-band nature of the tight-binding model (Fig.~\ref{figS2}\textbf{a}).
The oscillation amplitude of $P_{{z}}(\bm{q},t)$ vanishes at the $\Gamma$ point, while at the $k_1$ point, the oscillation amplitude reaches $1$, as shown in Fig.~\ref{figS2}\textbf{b}.

In the real system with Hamiltonian Eq.(1), the time evolution of the spin polarization $P_{{z}}(\bm{q},t)$ in the FBZ is simulated by numerical calculation. As a multi-band system, as shown in Fig.~\ref{figS2}\textbf{c}, the population of atoms oscillates between different bands after the quench. Thereby, $P_{{z}}(\bm{q},t)$ shows multi-frequency components. The spin polarizations $P_{{z}}(\bm{q},t)$ at $\Gamma$ and $k_1$ are shown in Fig.~\ref{figS2}\textbf{d}, and the frequencies are obtained by Fourier analysis. At the $\Gamma$ ($k_1$) point, the main frequencies are about 10kHz (1kHz,9.8kHz and 10.8kHz). So $P_{{z}}(\bm{q},t)$ in a multi-band system is formally written as
\begin{equation}
P_{{z}}(\bm{q},t)=\sum_{i=1}^{n}\{a_i(\bm{q})\cos[\omega_i(\bm{q})t]+b_i(\bm{q})\},\label{multi_bandP}
\end{equation}
where the oscillating frequency $\omega_i(\bm{q})=2\pi f_i(\bm{q})$ is the frequency difference of arbitrarily-chosen two bands. $a_i(\bm{q})$, $b_i(\bm{q})$ and $n$ are the oscillation amplitude, the global offset, and the number of frequencies, respectively. Furthermore, the interaction between the external environments and the 2D SO coupled system leads to damping~\cite{damp_Rabi}. Thereby, the spin polarization is formally written as
\begin{equation}
P_{z}(\bm{q},t)=\sum_{i=1}^{n}A_i(\bm{q})\cos[\omega_i(\bm{q})t]e^{-\frac{t}{t_1}}+B(\bm{q})e^{-\frac{t}{t_2}}+D(\bm{q}).\label{decay_multi_bandP}
\end{equation}
Here, the first term represents the consecutive deacy-oscillation with the frequency $\omega_i(\bm{q})$, the amplitude $A_i(\bm{q})$, and the damping rate $1/t_1$. The second term represents a pure decay with the amplitude $B(\bm{q})$ and damping rate $1/t_2$. The last term $D(\bm{q})$ is an offset. The evolution of the damped multi-frequency spin polarization $P_{{z}}$ at the $\Gamma$ and $k_1$ points are shown in Fig.\ref{figS2}\textbf{d}.

\begin{figure}[htbp]
\begin{center}
\includegraphics[width=0.5\linewidth]{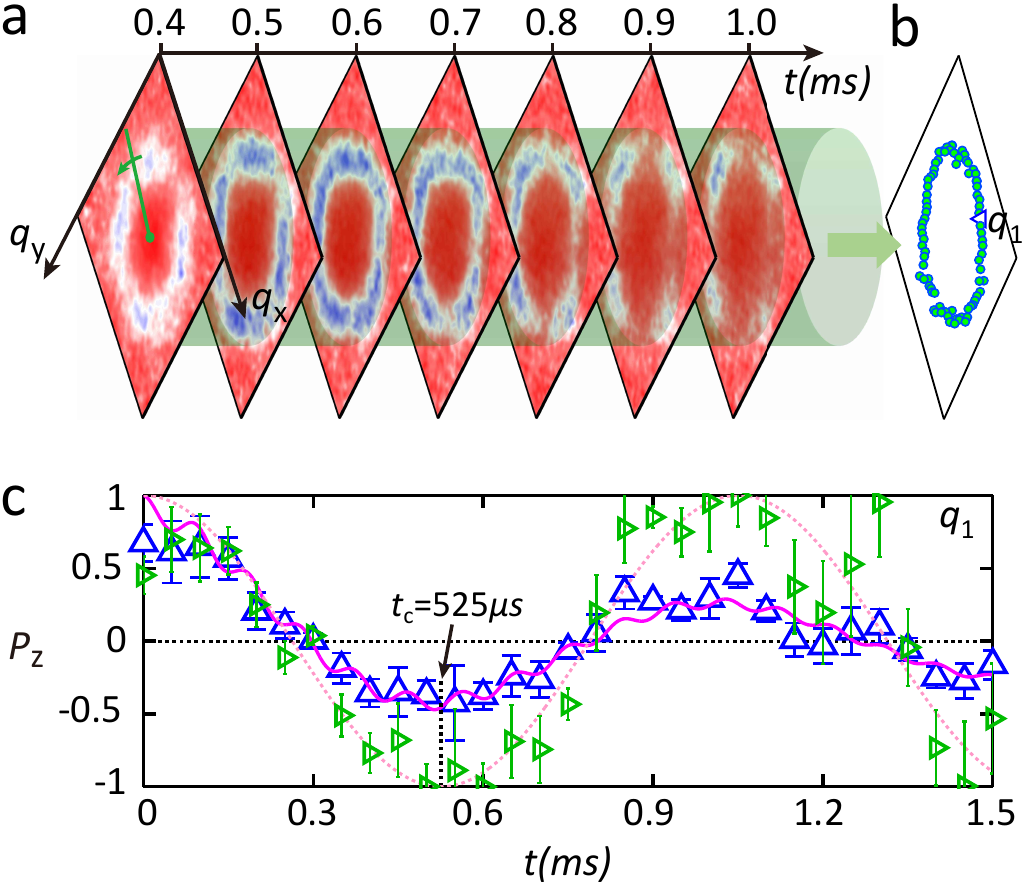}
\caption{
\textbf{Obtaining the trajectory $\mathcal{L_{-}}$ in the experiment.}
\textbf{a}, Finding the quasimomentum points on the trajectory $\mathcal{L_-}$. A green straight line is drawn from the quasimomentum point $(0,0)$ to the edge of FBZ. The quasimomentum points with minimal spin polarization ($k$-points) are obtained by the quasimomentum points on the green line throughout the evolution time.
\textbf{b}, The resulting $k$-points forms a {\lq\lq}ring" in the FBZ, which is obtained by rotating the green straight line.
\textbf{c}, The Raman-Rabi oscillations for the spin polarization at $q_1$,  whose locations are marked in \textbf{b}. The triangles with error bars are raw data. The magenta solid curve is the fitting with the fitting function Eq.~(\ref{decay_multi_bandP}). The green triangles with error bars are experimental data after removing the damping and higher band effects. The dashed pink curve is the fitting of the green triangle data with evolution of tight binding model (see the text). $t_{{c}}$ is the time that the spin polarization reaches its first local minimum.
}
\label{figS3}
\end{center}
\end{figure}

\subsection{Obtaining the trajectory $\mathcal{L_{-}}$ in the experiment}
In the following, we take the case with $V_0=4E_{{r}}, \Omega_0=1E_{{r}}$, and $\delta_{{f}}=0.2E_{{r}}$ as a example to illustrate the process of extracting the trajectory $\mathcal{L_{-}}$ from the experimental data. Firstly, the quasimomentum points on the trajectory $\mathcal{L_-}$ are identified. In order to fulfill this task, the FBZ is scanned around the $\Gamma$ point step by step throughout the whole evolution time, as shown in Fig.~\ref{figS3}\textbf{a}. In each step, the quasimomentum points with minimal spin polarization ($k$-points) are selected during the first oscillation period. $k$-points form a {\lq\lq}ring", as shown in Fig~\ref{figS3}\textbf{b}. Secondly, the time $t_{{c}}$ are determined by fitting $P_{{z}}$ at $k$-points using the fitting function Eq.~(\ref{decay_multi_bandP}). We intercept the $3$ frequencies from Eq.~(\ref{figS3}) to fit the experimental data. Fig.~\ref{figS3}\textbf{c} shows the evolution of the spin polarization at $q_1$. The high frequencies are about 10kHz corresponding to oscillation between $s$-band and $p$-band. The low frequencies $f_1$ are about several hundreds of Hertz ($952$Hz at $q_1$) corresponding to oscillation between two $s$-band. The decay time $t_1$ is about 1.2ms. According to the frequency difference between different bands, the information of the unitary evolution between the lowest two bands is extracted. Subsequently, the time $t_{{c}}$ are extracted by $t_{{c}}=1/(2f_1)$, i.e., $t_{{c}}$ is the time that the spin polarization reaches its first local minimum. Thus, the trajectory $\mathcal{L_-}$ is obtained.

\subsection{Hopf links for different detuning $\delta_f$ and Raman coupling $\Omega_0$ in the experiment}
\begin{figure}[htbp]
\begin{center}
\includegraphics[width=\linewidth]{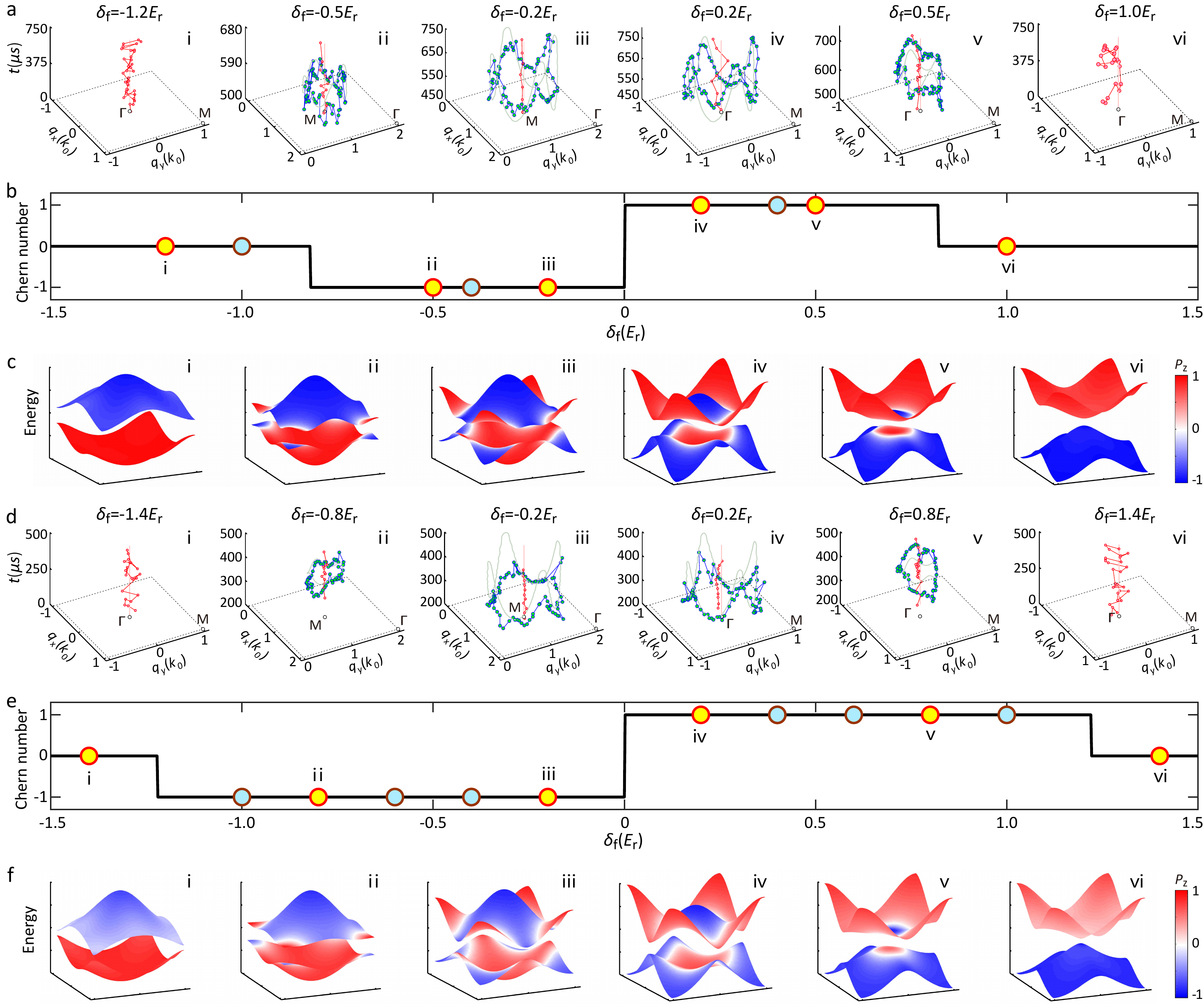}
\caption{
\textbf{Hopf links for different detunings $\delta_f$ and Raman couplings $\Omega_0$ in the experiment.}
\textbf{a}, The dynamical trajectories $\mathcal{L_+}$ (red dots) and $\mathcal{L_-}$ (blue dots) in $(\boldsymbol{q},t)$ space. The subfigures from {\romannumeral1} to {\romannumeral6} show the trajectories with $\Omega_{{0}}=1.0E_{{r}}$ and different detunings $\delta_f$.
The solid lines are from the theoretical calculation using tight binding model, and the dots are from the experiments. The relation between the parameters  for the continuous lattice  model $(V_0, \Omega_0)$ and the ones for the tight binding model $(t_0, t_{\text{so}})$ can be found in Ref.~\cite{uncover_topology}.
\textbf{b}, The change of the Chern number as the final detuning $\delta_{{f}}$.
\textbf{c}, Numerically calculated 2D SO coupling bands in equilibrium for the same parameters in \textbf{a}. The band inversions as shown in ii-iv are also a characteristic of topological bands, which have nonzero Chern numbers as indicated in \textbf{b}.
\textbf{d},\textbf{e},\textbf{f}, The same as in \textbf{a},\textbf{b},\textbf{c}, except that the Raman coupling strength is $\Omega_{{0}}=2E_{{r}}$.
}
\label{figS4}
\end{center}
\end{figure}

In this part, we present more data about Hopf links for different parameters, as shown in Fig.~\ref{figS4}.
In Fig.~\ref{figS4}\textbf{a}, we measure the trajectories $\mathcal{L}_{\pm}$ with the lattice depth $V_0=4.0E_r$, the Raman coupling strength $\Omega_0=1.0E_r$ and different detunings $\delta_f$ from $-1.2E_r$ to $1.0E_r$.
When $\delta_{{f}}<-0.82E_{{r}}$ ($\delta_{{f}}>0.82E_{{r}}$), the trajectory $\mathcal{L_-}$ vanishes, giving rise to the Hopf number $I_H=0$ and the Chern number ${C}=0$ (Fig.~\ref{figS4}\textbf{b}), which have trivial band structures (Fig.~\ref{figS4}\textbf{c}.{\romannumeral1} and {\romannumeral6}).
When $-0.82E_{{r}} < \delta_{{f}} < 0E_{{r}}$ ($0E_{{r}} < \delta_{{f}} < 0.82E_{{r}}$), the trajectories $\mathcal{L_-}$ and $\mathcal{L_+}$ are linked, giving rise to the Hopf number $I_H=-1$ $(+1)$ with the Chern number ${C}=-1$ $(+1)$ (Fig.~\ref{figS4}\textbf{b}), which have the topological nontrivial band structures (Fig.~\ref{figS4}\textbf{c}.{\romannumeral2}-{\romannumeral5}).
Other parameter with $V_{0}=4.0E_{{r}}, \Omega_{0}=2.0E_{{r}}$ and different detunings $\delta_{{f}}$ are also measured, as shown in Fig.~\ref{figS4}\textbf{d}, which has similar characterises.
To further demonstrate validity of the method, the two trajectories $\mathcal{L}_{\pm}$ calculated numerically are also drawn as the solid curves in Fig.~\ref{figS4}\textbf{a} and \textbf{d} from {\romannumeral1}-{\romannumeral6}. The numerical simulation agrees well with the experiment measurements.
The numerical calculation is executed by Hopf map Eq.(3), this is, $P_z({\bm{q},t})=-1 (+1)$ when $\theta=\pi (0)$. So the trajectory $\mathcal{L_-}$ and $\mathcal{L_+}$ is determined by $\left\{ h_z=0, t=\pi/{\left|h_f\right|} \right\}$ and $\left\{ h_{x}^{2}+h_{y}^{2} =0,t=[0,\infty) \right\}$, respectively.
The Chern number changing as the final detuning are shown in Fig.~\ref{figS4}\textbf{b} and \textbf{e}, which are consistent with the Chern number of the lowest band.

\subsection{Obtaining Hopf torus in the experiment}

\begin{figure}[htbp]
\begin{center}
\includegraphics[width=0.5\linewidth]{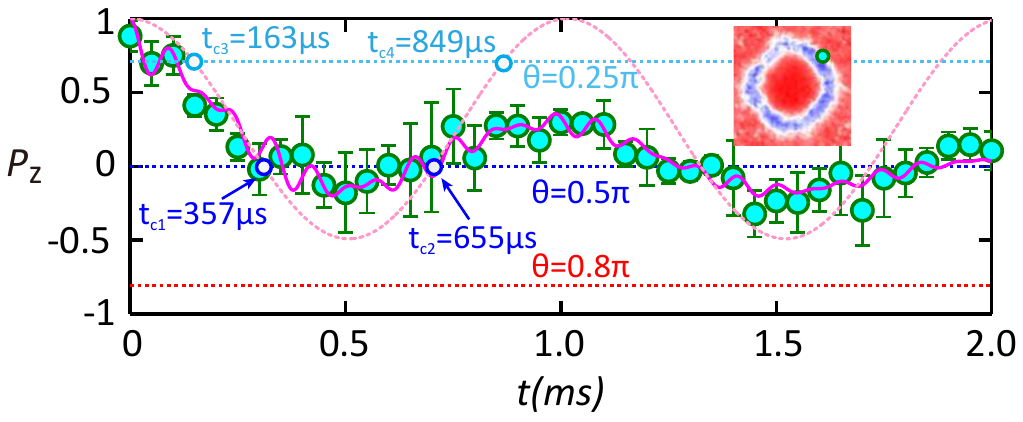}
\caption{
\textbf{Obtaining the Hopf torus in the experiment.}
The time evolution of the spin polarization at a certain momentum point $(q_x,q_y)=(0.48\pi, -0.5\pi)$, as marked in the inset. The circles with error bars are experimental data. The magenta solid curve is the fitting with Eq.~(\ref{decay_multi_bandP}). The dashed pink curve is the time evolution of the spin polarization corresponding to the two lowest bands in Eq.~(\ref{ideal_P}).
}
\label{figS5}
\end{center}
\end{figure}

According to the Hopf map in Eq.(3), the trajectories form a closed surface (the Hopf torus) in 2D quasimomentum plus time $(\bm{q},t)$ space when $P_z(\bm{q},t)=\cos(\theta)$ is satisfied. In order to obtain the Hopf torus in the experiment, $P_z(\bm{q},t)$ at each quasimomentum point in the FBZ is fitted by Eq.~(\ref{decay_multi_bandP}) and $P_{{z}}(\bm{q},t)$ between two lowest bands is extracted with the lowest fitting frequency. At each quasimomentum point, the characteristic time $t_c$ is extracted by the intersections of the curve $P_{{z}}(\bm{q},t)$ in the first oscillation period and the line $\cos(\theta)$. As a example, Fig.~\ref{figS5} draws Raman-Rabi oscillation at a certain point. There are two intersections between $P_{{z}}(\bm{q},t)$ and $\cos(0.5\pi)$ and the corresponding times are $t_{c1}=357\mu s$ and $t_{c2}=655\mu s$, respectively. There are two intersections between $P_{{z}}(\bm{q},t)$ and $\cos(0.25\pi)$ and the corresponding times are $t_{c3}=163\mu s$ and $t_{c4}=849\mu s$, respectively. However, there is no intersection between $P_{{z}}(\bm{q},t)$ and $\cos(0.8\pi)$. Thus, this quasimomentum point is removed from the FBZ for $\theta=0.8\pi$. This process is repeated for all the quasimomentum-time-dependent spin polarization: if the intersection exists between $P_{{z}}(\bm{q},t)$ and $\cos(\theta)$, the quasimomentum point and $t_c$ are retained; otherwise, they are removed. Finally, the Hopf tori are obtained as shown in Fig.4 in the main text.

\end{document}